\documentclass[aps,preprintnumbers,prl,twocolumn,superscriptaddress]{revtex4}

\usepackage{graphicx}

\usepackage{graphicx}
\usepackage{epstopdf}

\def\e{\epsilon}

\def\be{\begin{equation}}
\def\ee{\end{equation}}
\def\bea{\begin{eqnarray}}
\def\eea{\end{eqnarray}}

\def\bbuildrel#1_#2^#3{\mathrel{\mathop{\kern 0pt#1}\limits_{#2}^{#3}}}
\def\slash#1{\setbox0=\hbox{$#1$}#1\hskip-\wd0\dimen0=5pt\advance
       \dimen0 by-\ht0\advance\dimen0 by\dp0\lower0.5\dimen0\hbox
         to\wd0{\hss\sl/\/\hss}}

\newcommand{\gae}{\lower 2pt \hbox{$\, \buildrel {\scriptstyle >}\over {\scriptstyle
\sim}\,$}}
\newcommand{\lae}{\lower 2pt \hbox{$\, \buildrel {\scriptstyle <}\over {\scriptstyle
\sim}\,$}}

\def\issue(#1,#2,#3){{\bf #1}, #2 (#3)}
\def\opcit(#1){ {\em op. cit.}, #1}

\def\APP(#1,#2,#3){Acta Phys.\ Polon.\ \issue(#1,#2,#3)}
\def\ARNPS(#1,#2,#3){Ann.\ Rev.\ Nucl.\ Part.\ Sci.\ \issue(#1,#2,#3)}
\def\CPC(#1,#2,#3){Comp.\ Phys.\ Comm.\ \issue(#1,#2,#3)}
\def\CIP(#1,#2,#3){Comput.\ Phys.\ \issue(#1,#2,#3)}
\def\EPJC(#1,#2,#3){Eur.\ Phys.\ J.\ C\ \issue(#1,#2,#3)}
\def\EPJD(#1,#2,#3){Eur.\ Phys.\ J. Direct\ C\ \issue(#1,#2,#3)}
\def\IEEETNS(#1,#2,#3){IEEE Trans.\ Nucl.\ Sci.\ \issue(#1,#2,#3)}
\def\IJMP(#1,#2,#3){Int.\ J.\ Mod.\ Phys. \issue(#1,#2,#3)}
\def\JHEP(#1,#2,#3){J.\ High Energy Physics \issue(#1,#2,#3)}
\def\JPG(#1,#2,#3){J.\ Phys.\ G \issue(#1,#2,#3)}
\def\MPL(#1,#2,#3){Mod.\ Phys.\ Lett.\ \issue(#1,#2,#3)}
\def\NP(#1,#2,#3){Nucl.\ Phys.\ \issue(#1,#2,#3)}
\def\NIM(#1,#2,#3){Nucl.\ Instrum.\ Meth.\ \issue(#1,#2,#3)}
\def\PL(#1,#2,#3){Phys.\ Lett.\ \issue(#1,#2,#3)}
\def\PRD(#1,#2,#3){Phys.\ Rev.\ D \issue(#1,#2,#3)}
\def\PRL(#1,#2,#3){Phys.\ Rev.\ Lett.\ \issue(#1,#2,#3)}
\def\SJNP(#1,#2,#3){Sov.\ J. Nucl.\ Phys.\ \issue(#1,#2,#3)}
\def\ZPC(#1,#2,#3){Zeit.\ Phys.\ C \issue(#1,#2,#3)}

\def\simlt{\stackrel{<}{{}_\sim}}
\def\simgt{\stackrel{>}{{}_\sim}}

\begin{document}

\preprint{
\vbox{
\hbox{ANL-HEP-PR-09-10, EFI-09-01, MADPH-09-1529}
}}

\title{Family Non-universal $U(1)^\prime$ Gauge Symmetries and $b\rightarrow s$ Transitions}

\author{Vernon Barger}
\affiliation{Department of Physics, University of Wisconsin, Madison, WI 53706}
\author{Lisa Everett}
\affiliation{Department of Physics, University of Wisconsin, Madison, WI 53706}
\author{Jing Jiang}
\affiliation{Department of Physics, University of Wisconsin, Madison, WI 53706}
\author{Paul Langacker}
\affiliation{School of Natural Science, Institute for Advanced Study, 
             Einstein Drive, Princeton, NJ 08540}
\author{Tao Liu}
\affiliation{Enrico Fermi Institute, 
University of Chicago, 5640 S. Ellis Ave., Chicago, IL
60637}
 \author{Carlos E.M. Wagner}
\affiliation{Enrico Fermi Institute, 
University of Chicago, 5640 S. Ellis Ave., Chicago, IL
60637}
\affiliation{Kavli Institute for Cosmological Physics, 
University of Chicago, 5640 S. Ellis Ave., Chicago, IL
60637}
\affiliation{HEP Division, Argonne National Laboratory,
9700 Cass Ave., Argonne, IL 60439}

\begin{abstract}

We present a correlated analysis for the $\Delta B =1, 2$ processes which occur via $b\to s$ transitions within models with a family non-universal $U(1)^\prime$.  We take a model-independent approach, and only require family universal charges for the first and second generations and small fermion mixing angles.  The results of our analysis show that within this class of models, the anomalies in $B_s - \bar B_s$ mixing and the time-dependent  CP asymmetries of the penguin-dominated $B_d \to (\pi, \phi, \eta', \rho, \omega, f_0)K_S$ decays can be accommodated.

\end{abstract}

\maketitle

During the past several decades, great progress has been made in understanding CP-violating phenomena in the meson systems. With the advent of the B factories, it has been established that the observed CP violation in the $B_d$ and $K$ systems can be accommodated within the Standard Model (SM), sourced by the single phase of the Cabibbo-Kobayashi-Maskawa flavor mixing matrix \cite{1963}.  Despite the success of the SM picture of CP violation, it is known that additional sources of CP violation are needed in nature, for example to explain the origin of the baryon asymmetry of the universe.  Such additional phases, which arise in many models of new physics (NP) at the TeV scale, can result in competitive contributions to CP violation for processes in which the SM contribution enters at loop level,  such as in $b\rightarrow s$ transitions.  Indeed, recent measurements of a number of such observables exhibit discrepancies with SM predictions at the level of a few standard deviations, which may reveal the tantalizing possibility of physics beyond the SM.

The first relevant observable of this type is the $B_s-\bar B_s$ mixing phase, which is parametrized by the off-diagonal mixing matrix element  
\begin{eqnarray}
M_{12}^{B_s}=(M_{12}^{B_s})_{\rm SM} C_{B_s} e^{2 i \phi_{B_s}^{\rm NP}}.
\end{eqnarray}
In the SM, the modulus $C_{B_s}$ and phase $\phi_{B_s}^{\rm NP}$ are equal to one and zero, respectively.  A recent analysis~\cite{Bona:2008jn} finds that $\phi_{B_s}^{\rm NP}$ deviates by more than $3 \sigma$ from zero (see Table~\ref{table1}), by combining all the available experimental information on $B_s$  mixing, 
including the new tagged analysis of $B_s \to \psi \phi$ by CDF~\cite{Aaltonen:2007he} and D$\emptyset$~\cite{:2008fj}.   The discrepancy disfavors NP scenarios which obey minimal flavor violation (e.g., see~\cite{Tarantino:2008pb}), {\it i.e.}, with $\phi_{B_s}^{\rm NP}\approx 0$, and instead suggests NP which exhibits flavor violation in the $b\rightarrow s$ transitions.  
\begin{table}[th]
\begin{center}
\begin{tabular}{@{}ccc}
Observable  & $1 \sigma$ C.L.& $2 \sigma$ C.L.  \\
\hline
\hline
$\phi_{B_s}^{\rm NP} [^\circ]$  (S1)            & -19.9 $\pm$ 5.6 & [-30.45,-9.29] \\
    $\phi_{B_s}^{\rm NP} [^\circ]$                         (S2)       & -68.2 $\pm$ 4.9 & [-78.45,-58.2] \\
$C_{B_s}$                           & 1.07 $\pm$ 0.29 & [0.62,1.93] \\
\hline
\end{tabular}
\end{center}
\caption {The fit results for the $B_s - \bar B_s$ mixing parameters~\cite{Bona:2008jn}. The two solutions for $\phi_{B_s}^{\rm NP}$, denoted as ``S1'' and ``S2'', result from measurement ambiguities; see~\cite{Bona:2008jn} for details.}
\label{table1}
\end{table}

Another set of emblematic examples is the set of time-dependent CP asymmetries in the charmed and in the penguin-dominated charmless hadronic $b \to s \bar qq$ $(q = u, d, c, s)$ decays.  The SM predicts that many decays in this class, including $B_d\to \psi K_S$ and $B_d \to (\phi, \eta', \pi, \rho, \omega, f_0) K_S$, obey the relations (e.g., see~\cite{Barberio:2008fa}) 
\begin{eqnarray}
-\eta_{f_{CP}}{\mathcal S}_{f_{CP}} \simeq \sin 2 \beta,  \ \ \ \  {\mathcal C}_{f_{CP}} \simeq 0, 
\end{eqnarray}
in which $\eta_{f_{CP}}=\pm1$ is the CP eigenvalue for the final state $f_{CP}$ and $\beta \equiv \arg\left[-(V_{cd}V_{cb}^*)/(V_{td}V_{tb}^*)\right]$.
However, the central values of $\sin 2 \beta$ directly measured from the penguin-dominated modes are systematically below the SM prediction and the results obtained from measuring the charmed $B_d \to \psi K_S$ mode (see Table~\ref{table2}).
Given that $b\to s \bar cc$ is dominated by the SM tree-level amplitude, $\Delta {\mathcal S}_{f_{CP}}=-\eta_{f_{CP}}{\mathcal S}_{f_{CP}}+ \eta_{\psi K_S}{\mathcal S}_{\psi K_S}\neq 0$ may imply interesting NP in the penguin-dominated decay modes.

\begin{table}[t]
\begin{center}
\begin{tabular}{|c|c|c|}
  \hline
$f_{CP}$ & $-\eta_{CP} {\mathcal S}_{f_{CP}}$ (1$\sigma$ C.L.) & ${\mathcal C}_{f_{CP}}$(1$\sigma$ C.L.)   \\  \hline
$\psi K_S$  & $+0.672\pm0.024 $ & $+0.005\pm0.019 $   \\ \hline 
$\phi K_S$ & $+0.44^{+0.17}_{-0.18} $ & $-0.23\pm0.15 $  \\
$\eta^\prime K_S$ & $+0.59\pm0.07  $ & $-0.05\pm0.05$  \\  
$\pi K_S$ & $+0.57\pm0.17$ & $+0.01\pm0.10$  \\
$\rho K_S$ & $+0.63^{+0.17}_{-0.21}$ & $-0.01\pm0.20 $  \\ 
$\omega K_S$ & $+0.45\pm0.24$ & $-0.32\pm0.17 $  \\ 
$f_0 K_S$ & $+0.62^{+0.11}_{-0.13}$ & $0.10\pm0.13$ \\ \hline
\end{tabular}
\end{center}
\caption{World averages of the experimental results for the CP
  asymmetries in $B_d$ decays via $b\to\bar qq s$ transitions~\cite{Barberio:2008fa}.} \label{table2}
\end{table}

%
To account for these possible discrepancies in the $b\rightarrow s$ transitions, a number of NP scenarios have been studied (e.g., see~\cite{NPmodels}).  In this letter, we will study these anomalies within the NP scenarios with a family non-universal gauged $U(1)'$ symmetry (see also~\cite{Barger:2003hg}).  Additional $U(1)'$ gauge symmetries naturally exist in many well-motivated extensions of the SM (e.g., see~\cite{Langacker:2008yv}). The associated gauge charges for the SM fermions $\tilde \epsilon^{\psi_{L,R}}_{ij}$ can be diagonalized with a proper choice of gauge basis. Here $\psi$ represents the quarks and leptons, and $i, j$ are family indices.  For $U(1)^\prime$ models in which the gauge charges are family non-universal,
the $Z'$ couplings are affected by fermion mixings and are not diagonal in the mass basis. 
Non-trivial flavor-changing neutral current (FCNC) effects mediated by the $Z^\prime$ therefore are induced. The formalism for such effects is systematically developed in~\cite{Langacker:2000ju,BEJLLW}. Explicitly, the strong limits from $K-\bar K$ mixing and $\mu-e$ conversion exclude significantly non-universal couplings for the first two families for a TeV-scale $Z'$ with electroweak couplings, suggesting an approximate coupling structure in the basis of the mass eigenstates of 
$\psi_{L,R}$:
\begin{eqnarray}
B^{\psi_{L,R}} &\equiv& \frac{g_2 M_Z}{g_1 M_{Z'}}V_{\psi_{L,R}}\tilde \epsilon^{\psi_{L,R}} V_{\psi_{L,R}}^{\dagger} \nonumber \\
   & =& \begin{pmatrix} {  B^{\psi_{L,R}}_{11} &0&B^{\psi_{L,R}}_{13} \cr 0 &   B^{\psi_{L,R}}_{11}& B^{\psi_{L,R}}_{23} \cr B^{\psi_{L,R}*}_{13} & B^{\psi_{L,R}*}_{23} &  B^{\psi_{L,R}}_{33} }  \end{pmatrix},
\label{bcoup}
\end{eqnarray}
in which $V_{\psi_{L,R}}$ are the unitary matrices that diagonalize the fermion mass matrices.     
$B^{\psi_{L,R}}$ differs from the definitions in~\cite{Langacker:2000ju} by the factor $\frac{g_2 M_Z}{g_1 M_{Z'}}$ ($g_{1,2}$ denote the perturbative $Z$ and $Z'$ couplings). Hermiticity dictates that there are only two phases, associated with $B^{\psi_{L,R}}_{13}$ and $B^{\psi_{L,R}}_{23}$, respectively. Such a structure can be obtained by simply assuming $\tilde \e^{\psi_{L,R}}_{1}=\tilde \e^{\psi_{L,R}}_{2}$ and small fermion mixing angles~\cite{BEJLLW}. It demonstrates that non-universal $Z'$ effects can contribute to $\Delta B =1, 2$ processes via $b\to s$ transitions.  In this letter, we will neglect the effects of $B^{\psi_{L,R}}_{13}$, and assume no $Z-Z'$ mixing and  no $Z'$-mediated mixing between ordinary and exotic fermions 
which may also result in nontrivial FCNC effects (e.g., see~\cite{Langacker:2008yv}). 




The $Z'$ contributions to FCNC are at tree level, and hence they may be competitive even for small couplings.  Depending on the details of the $U(1)^\prime$ model, such $Z^\prime$ effects can result both in new FC operators and modified Wilson coefficients to the existing SM operators in the operator product expansion.  The details of this formalism for the $b\to s$ transitions will be presented in~\cite{BEJLLW}.  Here we summarize the necessary elements.  For $B_s-\bar B_s$ mixing and $b\to s \bar qq$, the modifications to the effective Hamiltonians  are respectively given by
\begin{eqnarray} 
&& \mathcal{H}_{eff}^{Z'}(B_s -\bar B_s) = 
\nonumber \\ &&
- \frac{G_F}{\sqrt{2}}  (\Delta C^{ B_s}_1Q^{B_s}_1 + 2\Delta \tilde C^{ B_s}_3 \tilde Q^{B_s}_3 +\Delta \tilde C^{ B_s}_1 \tilde Q^{B_s}_1 )+ h.c. \nonumber \\
 && {\cal H}_{\rm eff}^{Z'} (b \to s \bar q q) =  \nonumber \\
 && - \frac{G_F}{\sqrt{2}} V_{tb} V_{ts}^*(\Delta  C_3 Q^3 + \Delta C_5 Q^5 +  \Delta C_7 Q^7 +  \Delta C_9 Q^9  \nonumber \\&&
 + \Delta \tilde C_3 \tilde Q^3 + \Delta \tilde C_5 \tilde Q^5 +  \Delta \tilde C_7 \tilde Q^7 +  \Delta \tilde C_9 \tilde Q^9 ) +
\mbox{h.c.}, \label{117}
\end{eqnarray}
in which the $Q$s represent the SM operators~\cite{Buchalla:1995vs}, and the $\tilde Q$s are new operators introduced by the NP ($\tilde Q_3^{B_s}=(\bar s b)_{V+A} (\bar s b)_{V-A} $; the $Q$, $\tilde Q$ with the same superscript and subscript are related by chirality-flipping).  For simplicity, in this letter we will work in the limit that  $B_{bs}^L = B_{bs}^R$ for the quark sector, in which the nontrivial $Z'$ corrections to the Wilson coefficients are 
\begin{eqnarray}
 \Delta C^{B_s}_1& =& \Delta \tilde C^{B_s}_1 = \Delta \tilde C^{B_s}_3 =  - (B_{bs}^L)^2, \nonumber \\
\Delta C_{3} &=&   \Delta \tilde C_{5} = - \frac{2}{ V_{tb} V_{ts}^*}  B_{bs}^L B_{dd}^L,  \nonumber \\
\Delta \tilde C_{3} &=& \Delta C_{5}=- \frac{2}{3 V_{tb} V_{ts}^*}  B_{bs}^L \left(B_{u u}^R + 2 B_{dd}^R\right) , \nonumber \\
\Delta C_{7} &=& \Delta \tilde C_{9}  = -\frac{4}{3 V_{tb} V_{ts}^*}  B_{bs}^L\left(B_{u u}^R- B_{d d}^R \right). 
 \label{134} 
\end{eqnarray}    
Though $Z^\prime$-mediated effects can occur in both the QCD and electroweak (EW) penguins, we assume that they are mainly manifest in the EW ones, $i.e.$, $|\Delta C_{3,5}|\ll |\Delta C_7|$~\cite{Buras:2003dj,Barger:2003hg}.  In this limit, there are three independent parameters: the modulus of $B_{bs}^L$, its phase $\phi_{bs}^L$, and the real $B_{dd}^R(\simeq -B_{uu}^R/2)$.


\begin{figure}
\includegraphics[width= 0.70 \linewidth]{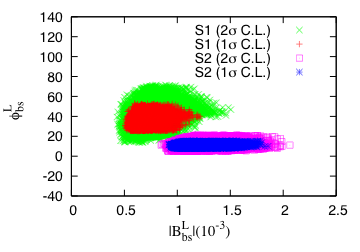}
\hspace{3.3cm}
\caption{Constraints on the $|B_{bs}^L|$ and $\phi_{bs}^L[^\circ]$ are presented.   Random values  for $C_{B_s}$ and $\phi_{B_s}^{\rm NP}$ from the physical regions (see Table~\ref{table1}) are mapped to the $|B_{bs}^L|-\phi_{bs}^L$ plane using Eq.~(\ref{20}). For the numerical work in this letter, a $25\%$ uncertainty (a typical value from non-perturbative effects) is universally assumed for the coefficients in the NP correction terms.}
\label{figure1}
\end{figure}

In contrast to the studies in~\cite{Barger:2003hg}, which are based on mode-by-mode analyses, we will take a correlated analysis for the $\Delta B =1, 2$ processes which occur via $b\to s$ transitions (for general discussions on the relevant calculations, see e.g.~\cite{Buchalla:1995vs}). We focus first on $B_s-\bar{B}_s$ mixing. With the renormalization scale chosen as the $b$-quark mass, $4.2$ GeV, 
the NP probes  $C_{B_s}$ and $\phi_{B_s}^{\rm NP}$ are given by
\begin{eqnarray}
C_{B_s} e^{2i\phi_{B_s}^{\rm NP}}& =&1 - 3.59 \times 10^5 ( \Delta C_1^{B_s}+ \Delta \tilde C_1^{B_s}) \nonumber \\ && + 2.04\times  10^6 \Delta \tilde C_3^{B_s}  \label{20}
\end{eqnarray}
which involves two of the three free parameters: $|B_{bs}^L|$ and $\phi_{bs}^L$. 
In Fig.~\ref{figure1}, their experimental constraints from $B_s-\bar{B}_s$ mixing are illustrated.  There are two separate shaded regions, corresponding to the two $\phi_{B_s}^{\rm NP}$ solutions (see Table~\ref{table1}). For each region,  the various colors of points specify different confidence levels (C.L.) of the relevant $C_{B_s}$ and $\phi_{B_s}^{\rm NP}$ values. To explain the discrepancy of the observed $B_s-\bar B_s$ mixing from the SM prediction, $|B_{bs}^L|$ is required to be $\sim 10^{-3}$.  This follows from Eq. (\ref{20}) and the fact that
$C_{B_s}$ does not deviate from its SM prediction significantly. The smallness of $|B_{bs}^L|$ can help satisfy the experimental constraints of the branching ratio ${\rm Br}(B_s\to \mu^+\mu^-)$ easily~\cite{BEJLLW}, and is generically consistent with our assumption of  small fermion mixing angles, since $B_{bs}^L$ is proportional to these mixing angles~\cite{BEJLLW}.

\begin{figure}
\includegraphics[width= 0.70 \linewidth]{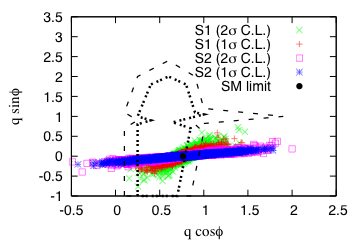}
\caption{It is illustrated how $B_{dd}^R$ is constrained through $q e^{i\phi}$. The points from the $|B_{bs}^L| - \phi_{bs}^L$ plane (see Fig.~\ref{figure1}) are randomly combined with 
scattered points of $B_{dd}^R$ ($10^{-3} \le |B_{dd}^R| \le 10^{-1}$) and then mapped to the $q\cos\phi -q\sin\phi$ plane according to Eq.~(\ref{21}). The colors of the points in this plane indicate the C.L. that their inverse images represent in Fig.~\ref{figure1}, and the two dashed lines specify the experimentally allowed ranges, due to $\chi^2$ fit of the $B\to\pi K$ (and $B\to\pi\pi$) data, at $1 \sigma$ and $90\%(\simeq 1.7\sigma)$ C.L., respectively~\cite{Fleischer:2008wb}. }
\label{figure2}
\end{figure}

A process of particular interest is the $B_d \to \pi K_S$ decay, which has received considerable interest in literatures (see e.g.~\cite{Buras:2003dj}-\cite{Ciuchini:2008eh}).  
In~\cite{Buras:2003dj} it is pointed out that a deviation of ${\cal S}_{\pi K_S}$ from its SM prediction can be understood as a modification of the ratio
$q e^{i\phi} = \frac {P} {T+C}$, 
in which $T$, $C$ and $P$ denote the color-allowed tree, color-suppressed tree, and EW penguin contributions in the decay amplitude, respectively. 
The non-universal $Z'$ modifies $qe^{i\phi}$ through 
\begin{eqnarray}
q e^{i\phi} &=& 0.76 (1 + 158.1 \Delta C_7 -102.4 \Delta \tilde C_9).  \label{21}
\end{eqnarray}
The constraints on $q e^{i\phi} $ from $\chi^2$ fit of the $B\to\pi K$ (and $B\to\pi\pi$) data have been obtained in~\cite{Fleischer:2008wb}. 
In Fig.~\ref{figure2},  we illustrate how $B_{dd}^R$ is constrained through $q e^{i\phi}$, using the parameter values of $|B_{bs}^L|$ and $\phi_{bs}^L$ obtained in Fig.~\ref{figure1} and Eq.~(\ref{21}). There are two distribution regions specified by different colors in this figure, again due to the two $\phi_{B_s}^{\rm NP}$ solutions.  According to Eq.~(\ref{21}), their right-up branches are related to positive $B_{dd}^R$ values while their left-bottom ones are related to negative $B_{dd}^R$ values.

\begin{figure}[ht]
\begin{center}
\includegraphics[width=0.23\textwidth]{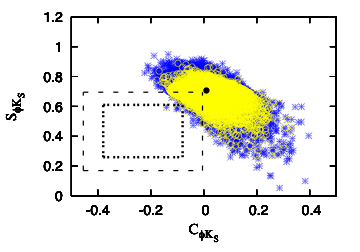}
\includegraphics[width=0.23\textwidth]{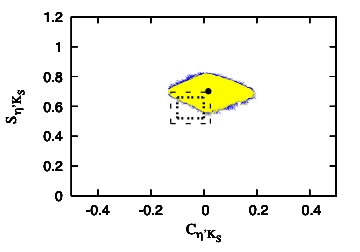}
\includegraphics[width=0.23\textwidth]{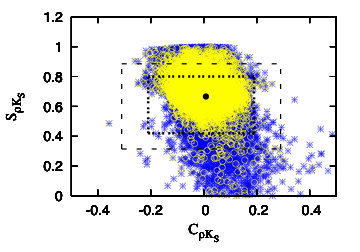}
\includegraphics[width=0.23\textwidth]{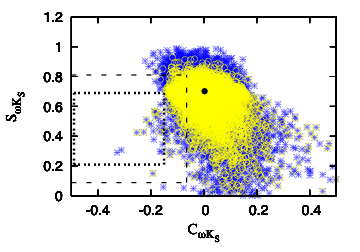}
\includegraphics[width=0.23\textwidth]{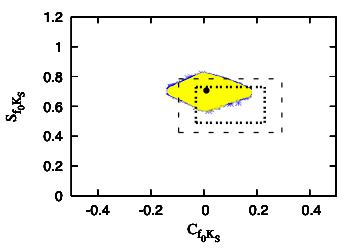}
\includegraphics[width=0.23\textwidth]{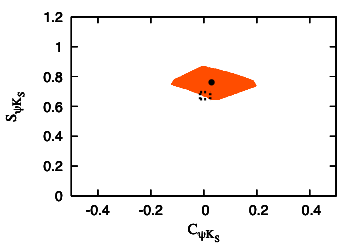}
\caption{
With the $B_{bs}^L$ and $B_{dd}^R$ values constrained by $B_s-\bar B_s$ mixing and $B_d\to \pi K_S$ decay, the NP contributions to  ${\mathcal C}_{(\phi, \eta', \rho, \omega, f_0)K_S}$ and ${\mathcal S}_{(\phi, \eta', \rho, \omega, f_0)K_S}$ are illustrated in the first five panels. The colors of the points specify the C.L. that their inverse image points represent in Fig.~\ref{figure1} and Fig.~\ref{figure2} (yellow denotes $1 \sigma$ C.L. in both and blue denotes $2 \sigma$ and $1.7 \sigma$ C.L., separately.). In the last panel the $CP$ asymmetries of the charmed $B_d\to \psi K_S$ decay are presented (with $|V_{ub}|=3.51\times 10^{-3}$ used in the SM calculation~\cite{CKMfitter08}). For each of these panels, the two boxes specify the 1$\sigma$ and $1.7\sigma$ experimentally allowed regions (except in the last one where only the $1 \sigma$ box is given), and the dark point denotes the SM limit.}
\label{figure3}
\end{center}
\end{figure}

To obtain acceptable values in both $B_s-\bar B_s$ mixing and $B_d \to \pi K_S$ from the SM predictions, the three free parameters in our scenario, $|B_{bs}^L|$, $\phi_{bs}^L$ and $B_{dd}^R$, must be constrained. The natural question is then whether the experimentally allowed values for these parameters can also be used to explain the ${\mathcal S}_{f_{CP}}$
anomalies in the remaining penguin-dominated $B_d \to (\phi, \eta', \rho, \omega, f_0)K_S$ decays.  
In Fig.~\ref{figure3}, we systematically illustrate the time-dependent $CP$ asymmetries in these modes, using the ``physical" parameter values obtained above. 
We universally assume a $15\%$ uncertainty in the SM calculations for each mode (as well as for the $B_d\to \pi K_S$ mode.). From the last panel where the NP effects are negligible, we see that this uncertainty is the least one necessary to simultaneously explain the experimental data of ${\mathcal C}_{\psi K_S}$, ${\mathcal S}_{\psi K_S}$ in the SM. The penguin-dominated modes have $0.5\sigma \sim 2 \sigma$ deviations for ${\mathcal C}_{f_{CP}}$, ${\mathcal S}_{f_{CP}}$, or both, except $B_d \to \rho K_S$. Because of interference effects between the $B_d-\bar B_d$ mixing phase and $\phi_{bs}^L$, the points in Fig.~\ref{figure3} are scattered away from the SM limits, and hence for each decay mode there are always some points lying in the $1\sigma$ region.

\begin{figure}
\includegraphics[width=0.23\textwidth]{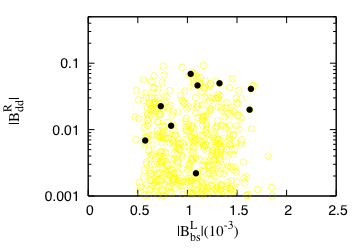}
\includegraphics[width=0.23\textwidth]{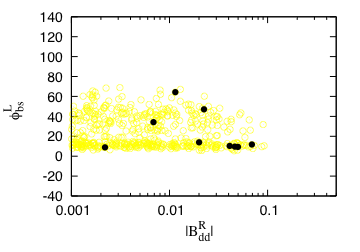}
\caption{The distributions of $|B_{bs}^L|$, $\phi_{bs}^L[^\circ]$ and $B_{dd}^R$ are illustrated, whose values are constrained by $B_s-\bar B_s$ mixing ($2 \sigma$ C.L.) and $\chi^2$ fit of the $B\to\pi K$ (and $B\to\pi\pi$) data ($1 \sigma$ C.L.), and then selected by  ${\mathcal C}_{(\phi, \eta', \rho, \omega, f_0)K_S}$, ${\mathcal S}_{(\phi, \eta', \rho, \omega, f_0)K_S}$ ($x \sigma$ C.L.), with the non-perturbative uncertainties in the SM and NP calculations assumed to be $15\%$ and $25\%$, respectively (because the hadronic matrix elements of the FC operators in the SM are better understood, compared to the new ones). Here $x = 1.7$ for yellow points and $x = 1.4$ for dark points.}
\label{figure4}
\end{figure}

To show that all of the constraints can be satisfied simultaneously, we carry out a correlated analysis among the $B_s-\bar B_s$ mixing and the $B_d \to (\pi, \phi, \eta', \rho, \omega, f_0)K_S$  $CP$ asymmetries. The allowed values for $|B_{bs}^L|$, $\phi_{bs}^L$ and $B_{dd}^R$ are illustrated in Fig.~\ref{figure4}. We see that indeed there exist parameter regions where all relevant anomalies 
can be explained at reasonable C.L.. The allowed $|B_{bs}^L|$ and $\phi_{bs}^L$ values can explain both solutions of $B_s - \bar B_s$ mixing, and the allowed $|B_{dd}^R|$ values are smaller than $0.08$. 
But, to get a better fit or a better agreement with the experimental data, 
$|B_{dd}^R| \simgt 10^{-2}$ is typically required. This effect is represented by the dark points in Fig.~\ref{figure4} which correspond the fitting of ${\mathcal C}_{(\phi, \eta', \rho, \omega, f_0)K_S}$ and ${\mathcal S}_{(\phi, \eta', \rho, \omega, f_0)K_S}$ with a higher standard. This effect can also be seen by requiring a smaller C.L. for the $\chi^2$ fit of $B\to\pi K$ (and $B\to\pi\pi$), which will be shown in details in~\cite{BEJLLW}. 
The favored parameter region for $|B_{dd}^R|$ is consistent with the assumption of trivial $Z'$ effects in QCD penguins and small fermion mixings
which requires $|B_{bs}^L| < |B_{dd}^L| \ll |B_{dd}^R|$. Because the value of $|B_{bs}^L|$ is favored to be $\sim 10^{-3}$, this relation can be easily accommodated. In addition this parameter region is of interest for collider detection. Given $(V_{d_R} \tilde \epsilon^{d_R} V_{d_R})_{11} \sim {\mathcal O}(1)$ it implies $\frac{g_1M_{Z'}}{g_2 M_{Z}} \sim 10-100$, a range approachable at the LHC for $g_2\simlt g_1$(e.g., see~\cite{Langacker:2008yv}). This fact is also important for the effective Hamiltonian in Eq. (\ref{117}) where the renormalization group running effect between the $Z'$ mass scale and the EW scale is neglected, which is justified only for a low scale $Z'$ boson.




Our results have been obtained in the limit of $B_{bs}^L = B_{bs}^R$. However, it is straightforward to extend the analysis to other limits, e.g.,  $\tilde \epsilon ^{\psi_L} (\tilde \epsilon ^{\psi_R}) \propto I$ while $\tilde \epsilon ^{\psi_R} (\tilde \epsilon ^{\psi_L})$  is only constrained by $\tilde \e^{\psi_{R}}_{1} (\tilde \e^{\psi_{L}}_{1}) =\tilde \e^{\psi_{R}}_{2} (\tilde \e^{\psi_{L}}_{1})$, or to generalize to the case where both of them are only constrained by this relation. Actually, similar results are found. We will present them elsewhere~\cite{BEJLLW}.

In conclusion, we have presented a correlated analysis for the $\Delta B =1, 2$ processes which occur via $b\to s$ transitions within NP models with a family non-universal $U(1)^\prime$.  In our model-independent approach, the main assumptions are generation-independent charges for the first two families and small fermion mixing angles.   We find that within this class of family non-universal $U(1)'$ models, the anomalies in $B_s - \bar B_s$ mixing and the time-dependent  CP asymmetries of the penguin-dominated $B_d \to (\pi, \phi, \eta', \rho, \omega, f_0)K_S$ can be accommodated.\\

\noindent{\it Acknowledgements} \\
We thank Cheng-Wei Chiang and Jonathan L.~Rosner for 
useful discussions. Work at ANL is supported in part by 
the U.S. Department of Energy (DOE), Div.~of HEP, Contract
DE-AC02-06CH11357.  Work at EFI is supported in part by the 
DOE through Grant No. DE-FG02- 90ER40560.  Work at the U.~Wisconsin, Madison is supported by the DOE through Grant No. DE-FG02-95ER40896 and the Wisconsin Alumni Research Foundation.  T.L. is also supported by the Fermi-McCormick Fellowship. The work of P.L. is supported by the IBM Einstein 
Fellowship and by NSF grant PHY-0503584.


\end{document}